\documentclass[aps,prl,twocolumn,showpacs]{revtex4}
\usepackage{epsfig}

\begin{document}

\title{Andreev reflection eigenvalue density in mesoscopic
    conductors.}  
\author{P. Samuelsson$^a$, W. Belzig$^b$ and Yu.~V.  Nazarov$^c$}
\affiliation{$^a$D\'epartement de Physique Theorique, Universit\'e
    de Gen\`eve, CH-1211 Gen\'eve 4, Switzerland. \\ $^b$ Department
    of Physics and Astronomy, Universit\"at Basel, Klingelbergstr.~82,
    4056 Basel, Switzerland.\\ $^c$ Department of Nanoscience, TU
    Delft, Lorentzweg 1, 2628 CJ Delft, The Netherlands.}

\begin{abstract} 
  The energy-dependent Andreev reflection eigenvalues determine the
  transport properties of normal-superconducting systems. We evaluate
  the eigenvalue density to get an insight into formation of resonant
  electron-hole transport channels.  The circuit-theory-like method
  developed can be applied to any generic mesoscopic conductor or
  combinations thereof. We present the results for experimentally
  relevant cases of a diffusive wire and a double tunnel junction.
\end{abstract}
\maketitle

Scattering theory provides a intuitive physical picture of transport
in mesoscopic conductors \cite{been97,blan00,nazarov:03}. A large
number of transport properties are characterized by the transmission
eigenvalues $\{T_n\}$ of the conductor, the eigenvalues of the matrix
product $tt^{\dagger}$, where $t$ is the matrix of the transmission
amplitudes and $n$ the channel index. Examples
are the Landauer formula for the conductance $G\sim\sum_nT_n$, the
shot noise power $P\sim \sum_nT_n(1-T_n)$ \cite{butt90} or, more
general, the full counting statistics $\sim \sum_n
\ln[1+T_n(\exp(i\chi)-1)]$ \cite{levitov:93}. Knowledge of the density
of the transmission eigenvalues for a given mesoscopic conductor
allows one to calculate the ensemble averaged transport properties. The
transmission eigenvalue density has been derived for generic
conductors such as diffusive wires \cite{wire,naz94a}, chaotic
cavities \cite{cavity}, double barrier junctions \cite{double,naz94b},
dirty tunnel barriers \cite{dirty}, or combinations thereof
\cite{naz94b}.

When a mesoscopic conductor is connected to a superconductor, the
electronic properties of the conductor are modified by the induced
proximity effect \cite{prox}. On a microscopic level, the proximity
effect results from Andreev reflection at the normal-superconducting
(NS) interface. In analogy to the normal conductor, the transport
properties in the most interesting regime, at energies well below the
superconducting gap, are naturally expressed in terms of the
\textit{Andreev reflection eigenvalues} $\{R_n\}$. The $\{R_n\}$ are
the eigenvalues of the matrix product
$s_{eh}^{\dagger}s_{eh}^{\phantom{\dagger}}$, where $s_{eh}$ is the
matrix of the Andreev reflection amplitudes. The conductance
\cite{nscond} $G_{NS} \sim \sum_nR_n$ and shot-noise power
\cite{nsnoise} $P_{NS}\sim \sum_nR_n(1-R_n)$ are the most investigated
quantities.

Due to the dephasing of electrons and holes with an energy difference
$2E$, the Andreev reflection eigenvalues depend on energy on the scale
of the Thouless energy $E_c$. Only at low energies $E \ll E_c$ are the
Andreev reflection eigenvalues simply related \cite{been92} to the
transmission eigenvalues as $R_n=T_n^2/(2-T_n)^2$, providing a direct
relation between the density of Andreev reflection eigenvalues (ARED)
and the density of transmission eigenvalues. To extend this relation
to finite energies, one would need not only the correlations between
transmission eigenvalues at different energies, but also the energy
dependent correlations between the transmission eigenvectors, i.e. the
statistics of the whole energy dependent scattering matrix. Apart from
being a complicated approach, this statistics is known only for a few
generic conductors \cite{been97}.  This makes it highly desirable to
find a way to directly calculate the ARED, without going via the
energy dependent statistics of the scattering properties of the normal
conductor.

In this paper we present a theoretical approach to the energy
dependent ARED for mesoscopic conductors.
It comes in the form of a circuit theory for quasiclassical Green's
functions \cite{naz94a,naz94b,naz97}. The theory is applied to a
diffusive wire junction, both with and without a tunnel barrier at the
NS-interface, and to a double tunnel-barrier junction. In all cases,
the induced proximity effect leads to an opening of electron-hole
channels, $R_n\approx 1$, at energies of the order of $E_c$. Our
results provide a unified and physically intuitive explanation for a
large number of theoretical and experimental results on current and
noise in NS-systems.
\begin{figure}[t]
\centerline{\psfig{figure=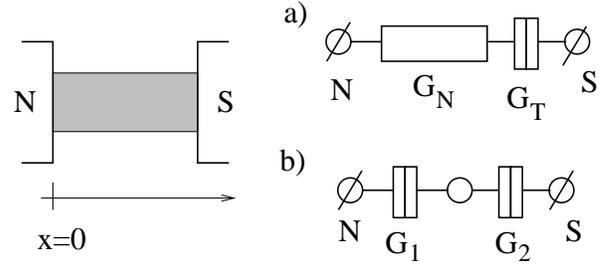,width=0.9\linewidth}}
\caption{Left: Schematic picture of a mesoscopic conductor
connected to a normal (N) and a superconducting (S) reservoir. Right:
The circuit theory representation of two mesoscopic conductors: a) a
diffusive wire in series with a tunnel barrier, b) two tunnel barriers
in series. The conductances of the respective circuit elements are
shown.}
\label{fig1}
\end{figure}

We consider a generic mesoscopic conductor with a large number of
transverse channels $N\gg1$ contacted to one normal and one
superconducting reservoir (see Fig. \ref{fig1}). An electron, incident
from the normal reservoir in channel $n$, has the amplitude
$(s_{eh})_{nm}$ to be backreflected as a hole in channel $m$. To
derive the ARED we follow Ref. \cite{naz94a} and define a
function in terms of the Andreev scattering amplitudes (depending on
energy) as
\begin{eqnarray}
H(E,\zeta)=\mbox{Tr}\left[\frac{s_{eh}^{\phantom{\dagger}}s_{eh}^{\dagger}}{1-\zeta^2
s_{eh}^{\phantom{\dagger}}s_{eh}^{\dagger}}\right]=\sum_n\frac{R_n}{1-\zeta^2 R_n}
\label{heq}
\end{eqnarray}
which gives all moments
Tr$\left[(s_{eh}^{\phantom{\dagger}}s_{eh}^{\dagger})^n\right]$ when
expanded in $\zeta$. From this function, the ARED $\rho(R,E)=\langle
\sum_n \delta(R-R_n)\rangle$ is given as
\begin{eqnarray}
\rho(R,E)&=&\frac{-1}{\pi R^2} \langle \mbox{Im}
\left[H\left(E,\zeta=\frac{1}{\sqrt{R+i0}}\right)\right]\rangle
\label{areddef} 
\end{eqnarray}
where $\langle...\rangle$ denotes average over impurity (or sample)
configurations. Next, the Andreev reflection amplitudes are related to
the retarded anomalous Gorkov Green's function $F^{\dagger R}({\bf
  x},{\bf x'})$, constructed from the scattering state solutions
\cite{Datta} to the Bogoliubov-de Gennes equation. Introducing
$F_{nm}^{\dagger R}(x,x')=\int
dydzdy'dz'\theta_n(y,z)\theta_n(y',z')F^{\dagger R}({\bf x},{\bf
  x'})$, with $\theta_n(y,z)$ the transverse wavefunction, we get
\begin{eqnarray}
\left(s_{eh}\right)_{nm}=-i\hbar \sqrt{v_nv_m} F_{nm}^{\dagger R}(x,x')
\label{seh}
\end{eqnarray}
with $F_{nm}^{\dagger R}(x,x')$ evaluated at the normal reservoir
($x=x'=0$) and where $v_n$ is the velocity in channel $n$. Using that
$F^{A}(x',x)=-[F^{\dagger R}(x,x')]^*$, the trace becomes
\begin{eqnarray}
\mbox{Tr}\left[s_{eh}^{\phantom{\dagger}}s_{eh}^{\dagger}\right]=-\hbar^2\sum_{n,m} v_nv_m
F_{nm}^{\dagger R}(x,x')F_{nm}^A(x',x).
\end{eqnarray}
Introducing the generalized quasiparticle current $I({\bf x},{\bf
x'})=\sum_n\hbar v_n \theta_n(y,z)\theta_n(y',z')\delta(x)\delta(x')$,
going from summation over modes to integration over transverse
coordinates, we can write
\begin{eqnarray}
\mbox{Tr}\left[s_{eh}^{\phantom{\dagger}}s_{eh}^{\dagger}\right]&=&\int d{\bf x}_1d{\bf
x}_2d{\bf x}_3d{\bf x}_4 I({\bf x}_1,{\bf x}_2) \nonumber \\
&\times&F^{R\dagger}({\bf x}_2,{\bf x}_3)I({\bf x}_3,{\bf x}_4)F^A({\bf x}_4,{\bf x}_1).
\label{gfexpr}
\end{eqnarray}
As pointed out in Ref. \cite{naz94a}, the right side of the expression
can be evaluated at an arbitrary cross-section of the normal
conductor. To obtain the traces of
$s_{eh}^{\phantom{\dagger}}s_{eh}^{\dagger}$ to all orders in Eq.
(\ref{heq}), we introduce a 4$\times$4-matrix Green's function in a
fictitious $\zeta$-dependent field, coupling the advanced and retarded
components as
\begin{eqnarray}
\check G({\bf x}_1,{\bf x}_2)&=&\check G^{0}({\bf x}_1,{\bf x}_2)+\int
d {\bf x}_3 d {\bf x}_4 \check G^{0}({\bf x}_1,{\bf x}_3) \nonumber \\
&\times& \zeta I({\bf x}_3,{\bf x}_4)\left[\check\tau_1+\check\tau_2\right]\check
G({\bf x}_4,{\bf x}_2).
\label{gfeq}
\end{eqnarray}
Here the unperturbed Green's function is
\begin{equation}
\check G^0=\left(\begin{array}{cc} \hat G^R & 0 \\ 0 & \hat G^A
\end{array} \right), \hspace{0.5cm} \hat
G^{R/A}=\left(\begin{array}{cc} G^{R/A} & F^{R/A} \\  F^{\dagger
R/A} & -G^{R/A} \end{array} \right),
\end{equation}
suppressing coordinate dependence, and $\check \tau_j$ are 
matrices with the only nonzero element being $(\check \tau_1)_{42}=1$
and $(\check \tau_2)_{13}=-1$. With these definitions, it follows from
the perturbation expansion in $\zeta$ that
\begin{equation}
\int d{\bf x}d{\bf x'} I({\bf x},{\bf x'})\mbox{tr}\left[\check \tau_1
\check G({\bf x},{\bf x'})\right]=\zeta H(E,\zeta),
\end{equation}
where the trace from now on is in the $4\times 4$ matrix
space. Starting from this expression, we can now apply the standard
techniques for impurity averaged quasi-classical Green's functions,
elaborated in detail for the circuit theory in
Refs. \cite{naz94b,naz97,bel03}. This gives the relation
\begin{equation}
\mbox{tr}\left[\check \tau_1 \check I(E,\zeta)\right]/G_Q=-i\zeta \langle
H(E,\zeta) \rangle
\label{circuit}
\end{equation}
where $\check I(E,\zeta)$ is the energy dependent matrix current
\cite{naz97} through the junction and $G_Q=2e^2/h$. As described in
detail in Ref. \cite{bel03}, the terms in Eq. (\ref{gfeq}) containing
$\zeta$ can be gauged away from the interior of the junction, giving
rise to renormalized boundary conditions for the normal reservoir
\begin{equation}
\check G_N=e^{i\zeta (\check \tau_1+\check \tau_2)}\check
G_{N0}e^{-i\zeta(\check \tau_1+\check \tau_2)}
\end{equation}
where $\hat G^{R/A}_{N0}=\pm \hat \sigma_z$ with $\hat \sigma_z$
denoting a Pauli-matrix in Nambu space. The boundary conditions in the
superconductor are not modified, i.e we have $\hat G_S^{R/A}= \hat
\sigma_x$. We can then, within the circuit theory for matrix Green's
functions Ref.~\cite{naz97}, calculate the matrix current $\check
I(E,\zeta)$. Knowing $\check I(E,\zeta)$, the ARED can be obtained
from Eq.  (\ref{areddef}) and (\ref{circuit}) as
\begin{equation}
\rho(R,E)=\frac{1}{G_Q}\frac{1}{\pi R^{3/2}}\mbox{Re}
  \left\{\mbox{tr}\left[\check \tau_1 \check
  I \left(E,\zeta=\frac{1}{\sqrt{R+i\eta}}\right) \right]\right\}.
\label{aredtot}
\end{equation}
This relation is the main technical result of the paper. 

To illustrate the approach, we first consider a diffusive wire with
conductance $G_N$ and length $L$ in good contact with a normal and a
superconducting reservoir. This system is described by the
quasiclassical Usadel equation \cite{usadel} $ D\partial_x \check G
\partial_x \check G=-iE [\hat\sigma_z,\check G]$, where $D$ is the
diffusion coefficient.  At low energies $E \ll E_c=\hbar D/L^2$, where the
proximity effect is fully developed, the density is obtained directly
from the transmission eigenvalue density of the wire \cite{wire},
giving
\begin{equation}
  \rho_{\text{diff}}=\frac{G_N}{4G_Q}\frac{1}{R\sqrt{1-R}}\,.
\end{equation}
This is just the same bimodal density as for the transmission
eigenvalues \cite{wire}, with a prefactor $1/2$.  In the opposite,
incoherent limit, $E \gg E_c$, the proximity effect is suppressed. It
was recently shown \cite{Belzig03b} that in this limit, not only the
current and the noise \cite{Nag}, but all moments of
the current, can be found by mapping the wire onto an effective normal
junction consisting of two diffusive wires in series. The same mapping
procedure can be used for the matrix Green's functions giving the
ARED. As a result the ARED is just given by the transmission
eigenvalue density of the effective junction (with conductance $G_N/2$
and the replacement $T\to R$, yielding $\rho(R,E \gg
E_c)=\rho_{\text{diff}}$, which is the same density as for low
energies.

\begin{figure}[t]
\centerline{\psfig{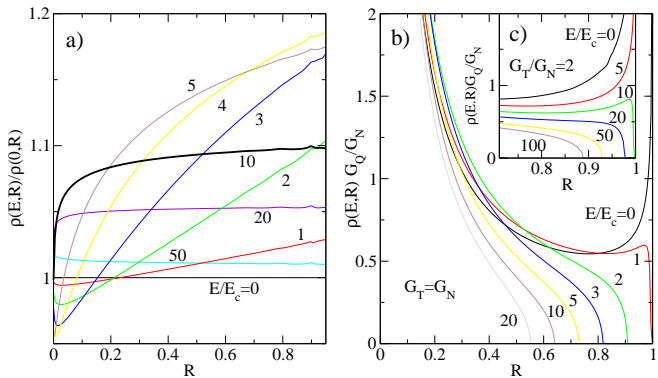}}
\caption{Energy-dependent Andreev reflection eigenvalue densities
  (ARED) of diffusive contacts. a) ARED for a diffusive wire ideally
  connected to a normal and a superconducting reservoir. Energies are
  measured in units of the Thouless energy $E_c=\hbar D/L^2$ and the
  ARED is normalized to the curve for $E=0$. At energies around $E_c$,
  resonant electron-holes channels, $R \approx 1$, open up. b) ARED
  with a contact resistance $G_T=G_N$ at the NS-interface. For finite
  energies the ARED has an upper bound; there is a gap in the density
  for eigenvalues above a maximum eigenvalue $R_{max}$. c) ARED with a
  contact resistance $G_T=2G_N$ at the NS-interface.  For $E\lesssim
  10 E_c$ all eigenvalues contribute, whereas a gap opens up for
  higher energies.}
\label{fig2}
\end{figure}

For energies of the order of $E_c$ the density differs
from $\rho_{\text{diff}}$. In this regime, the ARED was calculated
numerically, following the prescription given in
Ref.~\cite{Belzig}. The result is shown in Fig. \ref{fig2} (normalized
to the ARED at $E=0$). Increasing the energy from zero, the density of
open channels, $R\approx 1$, is enhanced in comparison to the density
of closed channels, $R\approx 0$, until $E\approx 5E_c$. For even
higher energies, the density becomes qualitatively similar to
$\rho_{\text{diff}}$ again, but with an overall enhancement
\cite{ared}. The induced proximity effect thus results in an opening
of electron-hole channels at energies of the order of $E_c$. This
provides a complementary picture to the explanation \cite{prox} that
an interplay of suppressed density of states and long-range
pair-correlations leads to a modification of the transport properties
of energies around $E_c$.

In particular, the opening of additional electron-hole channels
provides a simple explanation to the reentrant conductance peak
\cite{NazStoof}. Moreover, the qualitative difference between the
regimes $E\lesssim 5E_c$ and $E\gtrsim 5E_c$ explains the suppression
of the effective charge $q_{\text{eff}}(E)=(4/3)e \langle\sum_n
R_n(1-R_n)\rangle/\langle \sum_n R_n\rangle$ predicted theoretically
in \cite{Belzig} and found experimentally in
\cite{Reulet:03}. The overall enhancement of the ARED for
$E\gtrsim 5E_c$ cancels from the effective charge and explains why
$q_{\text{eff}}=2e$ for $E\gtrsim 5E_c$. In the regime of strong
proximity $E\lesssim 5E_c$ the enhancement of the density of open
channels results, due to the factor $1-R_n$, in a lowering of
$q_{\text{eff}}$.

In the presence of a tunnel barrier ($I$) at the NS-interface (see
Fig.\ \ref{fig1}a), it was found experimentally that due to the
proximity effect the zero-voltage conductance is strongly enhanced
\cite{NISexp} compared to what was expected from classical addition of
resistances of the wire and the NIS-interface. This was explained in
Ref.~\cite{reflectionless} as \textit{reflectionless tunneling}.
Backscattering by disorder in the wire enhances the chances for
Andreev reflection and thus increases the conductance.
Quantitatively, the ARED for a tunnel barrier with conductance $G_T$
at the NS-interface can be calculated numerically in the same way as
without the barrier. In Fig. \ref{fig2}b we present the ARED for
different energies in the case $G_T=G_N$. For low energies $E \ll E_c$
there is a finite fraction of completely open channels, $R=1$.
Increasing the energy, the whole density shifts towards lower
eigenvalues and the open channels start to close, i.e. there is a gap
in the density above a maximum reflection eigenvalue
$R_{\text{max}}<1$. A similar behavior is seen for $G_N<G_T$, except
that the opening of the gap occurs for a finite energy. In the
example, shown in Fig.~\ref{fig2}c for $G_T=2G_N$, the gap opens for
$E\gtrsim 10 E_c$.  Note, that in the limit $G_N\ll G_T$ where the
tunnel barrier can be neglected and there is a fraction of open
channels for all energies.  For a dominating tunnel barrier
resistance, $G_T<G_N$, there are no completely open channels even for
$E=0$. In the limit $G_T \ll G_N$ the shot-noise was recently observed
experimentally \cite{Lefloch:03} and studied theoretically
\cite{nsnoisetheory}. A doubling of the full Schottky noise was found
which can be understood by our results as a consequence of a
concentration of the ARED at small $R$ for $G_T \ll G_N$.

\begin{figure}[t]
\centerline{\psfig{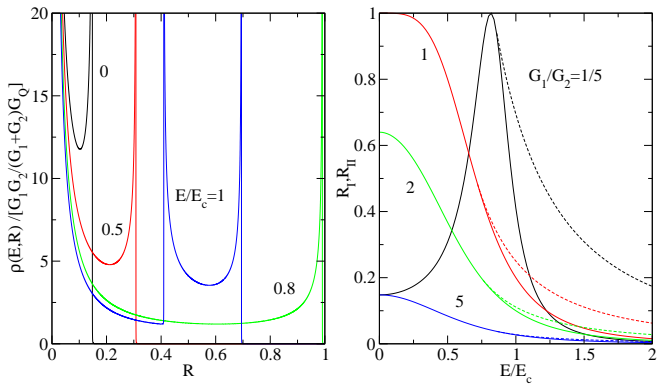}}
\caption{Left panel: ARED for a double-barrier junction with
$G_2=5G_1$ for different energies $E/E_c$. Right panel: Limiting
reflection eigenvalues $R_I$ (solid) and $R_{II}>R_I$ (dashed) as a
function of energy for different conductance ratios $G_1/G_2$}
\label{fig3}
\end{figure}

As a second example, we investigate the ARED in a double
tunnel-barrier junction (see Fig. \ref{fig1}b), a diffusive conductor
with negligible resistance connected to one superconducting and one
normal reservoir via tunnel barriers with conductance $G_1$ (to $N$)
and $G_2$ (to $S$).  The relevant energy scale for the proximity
effect is now $E_{c}=(G_1+G_2)\delta/(4G_Q)$, the inverse escape time
from the diffusive region.  Here $\delta$ is the level spacing in the
normal region.  For this system the ARED can be obtained analytically.
The solution for the Green's function of the diffusive conductor for
arbitrary energies is given in Refs.~\cite{Sam}.  Eq. (\ref{aredtot})
gives after some algebra the ARED $\rho\equiv\rho(R,E)$
\begin{eqnarray}
\rho&=&\frac{G_1^2G_2^2}{2\pi G_Q R^{5/4}}\left\{ 
\begin{array}{cc} \frac{1}{p\sqrt{p-\alpha \sqrt{R}}},& 0<R<R_I \\
\frac{1}{q\sqrt{-q-\alpha \sqrt{R}}}, & R_I<R<R_{II}\end{array}
\right. 
\label{NINIS}
\end{eqnarray}
where $p=(4G_1^2G_2^2-R\beta^2)^{1/2}$,
$q=(R(\alpha^2+\beta^2)-4G_1^2G_2^2)^{1/2}$
$\alpha=G_1^2+G_2^2-(E/E_{c})^2(G_1+G_2)^2$ and
$\beta=2G_1(G_1+G_2)E/E_{c}$. The limiting eigenvalues are
$R_I=4G_1^2G_2^2/(\alpha^2+\beta^2)$ and $R_{II}=4G_1^2G_2^2/\beta^2$
(see Fig. \ref{fig3}), noting that the region $R_{I}<R<R_{II}$ in the density in
Eq. (\ref{NINIS}) is present only for $\alpha<0$.

The ARED is plotted in Fig. \ref{fig3} for $G_2=5G_1$, where the
various properties of the ARED are clearly seen. For low energies
$E<E_c(G_1^2+G_2^2)^{1/2}/(G_1+G_2)$, i.e. $\alpha>0$, the ARED is
limited by an upper bound $R_I$ and is bimodal, see the upper term Eq.
(\ref{NINIS}). In the limit $E\ll E_c$ the ARED reduces to the one
obtained from the known \cite{naz94b} transmission eigenvalue density
for the normal double tunnel-barrier junction. For higher energies,
$E>E_c(G_1^2+G_2^2)^{1/2}/(G_1+G_2)$, the lower term in Eq.
(\ref{NINIS}) contributes as well. This leads to a qualitatively new
behavior of the ARED. The peak at the upper bound of the ARED
\textit{splits} and an extra bimodal band of Andreev reflection
eigenvalues between $R_I$ and $R_{II}$ emerges. Such a peculiar
distribution with three singular peaks has to our knowledge not been
observed before. 
We attribute this to a rigidity in position of the electron-hole
resonaces in double barrier junctions, most pronounced for $G_1\ll
G_2$. Upon changing the impurity configuration, each resonance
fluctuates in energy between values related to the conductances $G_1$
and $G_2$. As a consequence, for energies below the lowest resonance,
$\alpha>0$, the reflection eigenvalues take on values between $0$ and
$R_I$ and the density is bimodal. For higher energies, $\alpha<0$, the
peaks at $0$ and $R_{I}$ persist, but there is an additional peak at
the maximum reflection eigenvalue $R_{II}$ on resonance. In a junction
dominated by impurity scattering, this rigidity effect is washed out.

There are two physically different regimes for the ARED, depending on
the relation between the conductances $G_1$ and $G_2$. On one hand,
for dominating coupling of the diffusive conductor to the normal
reservoir, $G_1>G_2$, the value of $R_{\text{max}}\equiv
\mbox{max}(R_I,R_{II})$ never reaches unity, i.e. we have no
completely open channels. This situation resembles the NIS-junction
investigated above, i.e. the tunnel barrier to the normal reservoir
plays the same role as the diffusive wire.
On the other hand, for dominating coupling to the superconductor,
$G_2>G_1$, there is always an energy,
$E_{c}\sqrt{(G_2-G_1)/(G_2+G_1)}$, for which $R_{\text{max}}$ reaches
unity. Of particular interest is the limit $G_2 \gg G_1$. In this
limit, a proximity gap of magnitude $E_c$ is induced in the diffusive
conductor. Only at energies very close to $E_c$, which is exactly at
the edge of the induced gap, there are now open channels.

In conclusion, we have provided a theory for the Andreev reflection
eigenvalue density in mesoscopic normal-superconducting junctions. We
find quite generally that the proximity effect leads to an opening up
of resonant electron-hole channels at energies of the order of the
Thouless energy $E_c$. This was exemplified by studying a diffusive
wire and a double tunnel-barrier junction.

The work was supported by the Swiss network MaNEP (P.S.), the NCCR
Nanoscience and the Swiss NSF (W.~B.).


\begin{thebibliography}{999}
\bibitem{ared} 
The ARED is defined such that $G(E)=2 G_Q \int dR
\rho(E,R) R$ is the spectral conductance.
\end{thebibliography}
\end{document}